\documentstyle[12pt,epsfig]{article}
\begin{document}
\begin{center}
\vspace{1.5in}
{\LARGE
New proton and neutron magic numbers in neutron rich nuclei }
\end{center}
\vspace{.4in}
\begin{center}
{\bf Afsar Abbas}\\
\vspace{.1in}
Institute of Physics\\ 
Bhubaneshwar-751005, India\\
\vspace{,1in}
email: afsar@iopb.res.in
\end{center}
\vspace{1.5in}
\begin{center}
{\bf Abstract}
\end{center}
\vspace{.3in}

It is now known that in neutron rich nuclei, old magic numbers disappear 
and new ones appear. Single nucleon and double nucleon separation 
energies are plotted here in all possible manner.Using this data it is 
shown here for the first time that nuclei with pair of proton number Z 
and neutron number N ( Z,N ) : (6,12), (8,16), (10,20), (11,22) and 
(12,24) exhibit exceptional stability or magicity. As such these magic 
numbers appear in pairs. This correlation is shown here to be indicative of 
predominance of tritons in the ground state of these neutron rich nuclei. 
Thus ${^{30}_{10} Ne_{20}}$ has the structure of 10 $^{3}_{1} H_{2}$ 
in the ground state. This confirms the prediction of the author that 
triton clustering in neutron rich nuclei as a new phenomenon.

\newpage

The study of neutron rich nuclei with the help of RI beams has been
proving to be full of surprizes. First the halo structure cropped up and
now the realization that in the neutron rich nuclei the standard magic
numbers may disappear and new ones may arise in an unanticipated
manner [1,2,3,4]. In this letter the focus is the study of
these new magicities in neutron rich nuclei. Single nucleon and double
nucleon separation energies are analyzed in the standard manner and 
also in a slightly different manner as discussed below. 
Evidence for new magic numbers will be discussed. Interestingly 
proton and neutron numbers are found to be uniquely and consistently
correlated in these new magiciies. It shall be pointed out that the only 
possible way to understand this correlation is through the effect of
triton clustering in neutron rich nuclei. It will be emphasized that
triton clustering - a new effect predicted by the author [5], is for real.

To discuss magicity one normally plots separation energies
$S_{1n}$  and $S_{2n}$ as a function of neutron number $N$ for a
particular proton number $Z$ or plot $S_{1p}$  and $S_{2p}$ as a
function of $Z$ for a particular $N$.  
It is very common to plot $S_{2n}$ and look for kinks
as evidence for magicity [3,4]. In addition to plotting the separation 
energies in the above standard manner,
we plot the separation energies a little differently.  We plot
$S_{1n}$ and $S_{2n}$ as a function of $Z$ for a particular $N$.
The same with  $S_{1p}$ and $S_{2p}$ as a function of N for a particular
Z. We shall find here that these new plots give useful and 
distinguishable information regarding stabilty and magicity in these
light nuclei.

We plot separation energies for neutron number N and for proton number Z
fixed separately at 4, 6, 8, 10, 11, 12, 16, 20, 22 and 24.
For example Fig 1 and Fig 2 are respectively plots of  
 $S_{1n}$ and $S_{2n}$  for fixed N=
 4, 6, 8, 10, 11, 12, 16, 20, 22 and 24 plotted as a function of Z.
Figs 3 and 4 are respectively plots of 
$S_{1n}$ and $S_{2n}$ for fixed Z= 4, 6, 8, 10, 11, 12, 16, 20, 22 and 24
as a function of N. 
Figs 5 and 6 plot  $S_{1p}$ and $S_{2p}$ respectively for N fixed at 
the above set as a function of Z and Figs. 7 and 8 are
$S_{1p}$ and $S_{2p}$  plotted respectively for Z fixed at the above set 
for different N. So what we have plotted are N fixed ( isotones )
for $S_{1n}$, $S_{2n}$, $S_{1p}$ and $S_{2p}$ 
all as functions of Z and also for Z fixed ( isotopes )
$S_{1n}$, $S_{2n}$, $S_{1p}$ and $S_{2p}$ all as functions of Z.

In every case we notice the extra stabilty for all the cases wherein N=Z. 
This extra stability for even-even N=Z nuclei is attributed to the 
significance of $\alpha$ clusters. So for ${^{20}_{10} Ne_{10}}$
ground state it is the 5-$\alpha$ configuration which is believed to be 
important and thereby providing extra stabilty.

The reader's attention is drawn to the extra-ordinary stabilty manifested 
by the plotted data for the proton and neutron pairs (Z,N): 
(6,12), (8,16), (10,20), (11,22) and (12,24). The stabilty at these pair
of numbers is sometimes as prominent as that at the N=Z pair. In fact 
the Z,N pair (10,20) stands out as the best example of this.  
Hence it is clear that the separation energy data very clearly shows 
that there are new magicities present in the neutron rich sector for the 
pair (Z,N) where N=2Z.

What is the significance of this extraordianry stabilty or magicity for
the nuclei ${^{3Z}_Z} A _{2Z}$? We already know that for the even-even Z=N 
cases it is the significance of $\alpha$ clustering for the ground state 
of these nuclei which explains this extra stabilty.
Quite clearly the only way we can explain the extra magicity for these
N=2Z nuclei is by invoking the significance of triton clustering
in the ground state of these netron rich nuclei. 
Thus ${^{30}_{10} Ne _{20}}$ has significant mixture of the 
configuration 10 $^{3}_{1} H_{2}$ in the ground state.
It is these tritonic clusters which give the extra stabilty to these 
nuclei thereby provoding us whith these unique new
sets of magic numbers.

This tenedency of triton clustering in neutron rich nuclei was already 
predicted by the author a few years ago [5]. 
The present paper shows that there is indeed 
strong empirical evidence in support of the author's predicion [5].
If fact this may be treated as "smoking gun" kind of evidence in 
support of significane of triton clusters in the ground state of
${^{3Z}_Z} A _{2Z}$ nuclei.

One may ask to why the stabilty for a particular set is not manifesting 
itself in all the plots. One reason is that due to the limitation of the
span of individual isotones and individual isotopes over the 
corresponding (Z,N) sets may not be rich enough.
So it seems that neither the very low mass nor the high mass nuclei
would provide discriminatory evidence in this regard. Intermediate mass 
nuclei like ${^{30}_{10} Ne _{20}}$ tend to show up this 
effect most prominently for the isotonic data. 
However there may be other pairs like (4,8) 
where though somewhat weaker there does appear to be 
manifestation of possible magic pairs.
In this regard significance of the
presence of different clusters like $\alpha$ and/or tritons for the ground 
state of a particular nucleus 
and also whether it is proton/protons or neutron/neutrons being 
pulled out will manifest itself in the structure of the data. 
This point is at present under investigation by the author.

Once the triton clustering aspect of nucleus ${^{3Z}_Z} A _{2Z}$ is
understood then extra stabilities in the intermediate cases
like N=20 stabilty in $S_{2n}$ data for Z=16 can be understood as 
tendency of triton clustering to bring it about. One should be able to 
understand theses extra stabilities as mixing in the shell model 
configuration of $\alpha$ - clustering and t-clustering configurations. 
Similarly the helion ("h")  ${^{3}_{2} He_{1}}$~~ 
configuration may manifest itself especially in
odd-odd N=Z nuclei. One alreday knows that h-t configuration mixes 
considerably with $\alpha$-n-p configuration to give a good 
understanding of the ground state of  ${^{6}_{3} Li_{3}}$ .
Similar must be the case for the Z=12 stabilty of N=22 case of 
$S_{2n}$ where all - $\alpha$-, t- and h- clustering 
may all be playing a role. 
Such cases  have to be sorted and worked out in the future.
Also there would be other explicit manifestations of triton clustering
in data and which should be looked for. Here the purpose has been to bring 
out a few clear cut cases of the same so that there can be no doubt as
to the correctness of the idea [5].

In short, here the aim is to understand as to which new magic 
numbers are arising in the neutron rich nuclei. With this aim
single nucleon and double
nucleon separation energies are plotted here in all possible manner.
It is shown here that nuclei with
pair of proton number Z and neutron number N (Z,N) : (6,12), (8,16),
(10,20), (11,22) and (12,24) ( and possibly (4,8) ) 
exhibit exceptional stability or magicity.
This correlation is shown
to prove the significance of tritons in the ground state of
these neutron rich nuclei. Thus ${^{30}_{10} Ne_{20}}$ has the structure
of 10 $^{3}_{1} H_{2}$ in the ground state. 
This confirms the prediction of the
author on the tendency of triton clustering in neutron rich nuclei [5].
Whole new questions shall arise from this new empirical fact.

\vspace{.1in}

Acknowledgement: I would like to thank Dr P Arumugam for the figures.

\vspace{.2in}

{\bf References} 

\vspace{.2in}

1. I. Tanihata, Nucl. Phys. {\bf A682}, (2001) 114c

\vspace{.1in}

2. R. Kanungo, I. Tanihata and A. Ozawa, 
Phys. Lett.  {\bf B 528} (2002) 58

\vspace{.1in}

3. Z. Dlouhy, D. Baiborodin, J. Mrdzek and G.
Thiamova, Nucl. Phys. {\bf A722} (2003) 36c

\vspace{.1in}

4. M. Thoennessen, T. Baumann, J. Enders, N. H.
Frank, P. Heckman, J. P. Seitz and E. Tryggestad, Nucl. Phys.
{\bf 26} (2003) 61c

\vspace{.1in}

5. A. Abbas, Mod. Phys. Lett. {\bf A 16} (2001) 755

\newpage

\begin{figure}
\caption{Single neutron separation energy as a function of Z for fixed N}
\epsfclipon
\epsfxsize=0.99\textwidth
\epsfbox{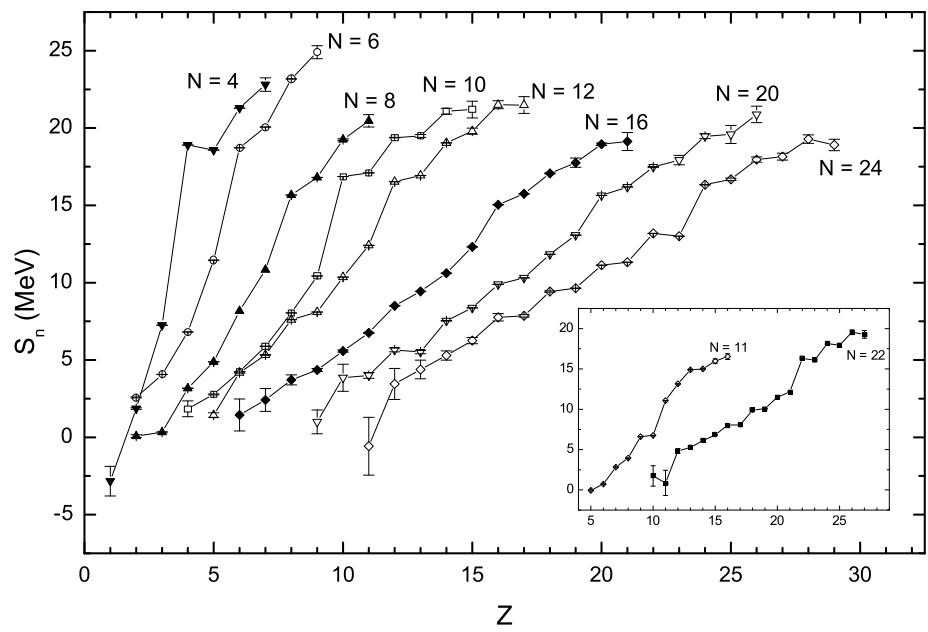}
\end{figure}

\newpage

\begin{figure}
\caption{Double neutron separation energy as a function of Z for fixed N}
\epsfclipon
\epsfxsize=0.99\textwidth
\epsfbox{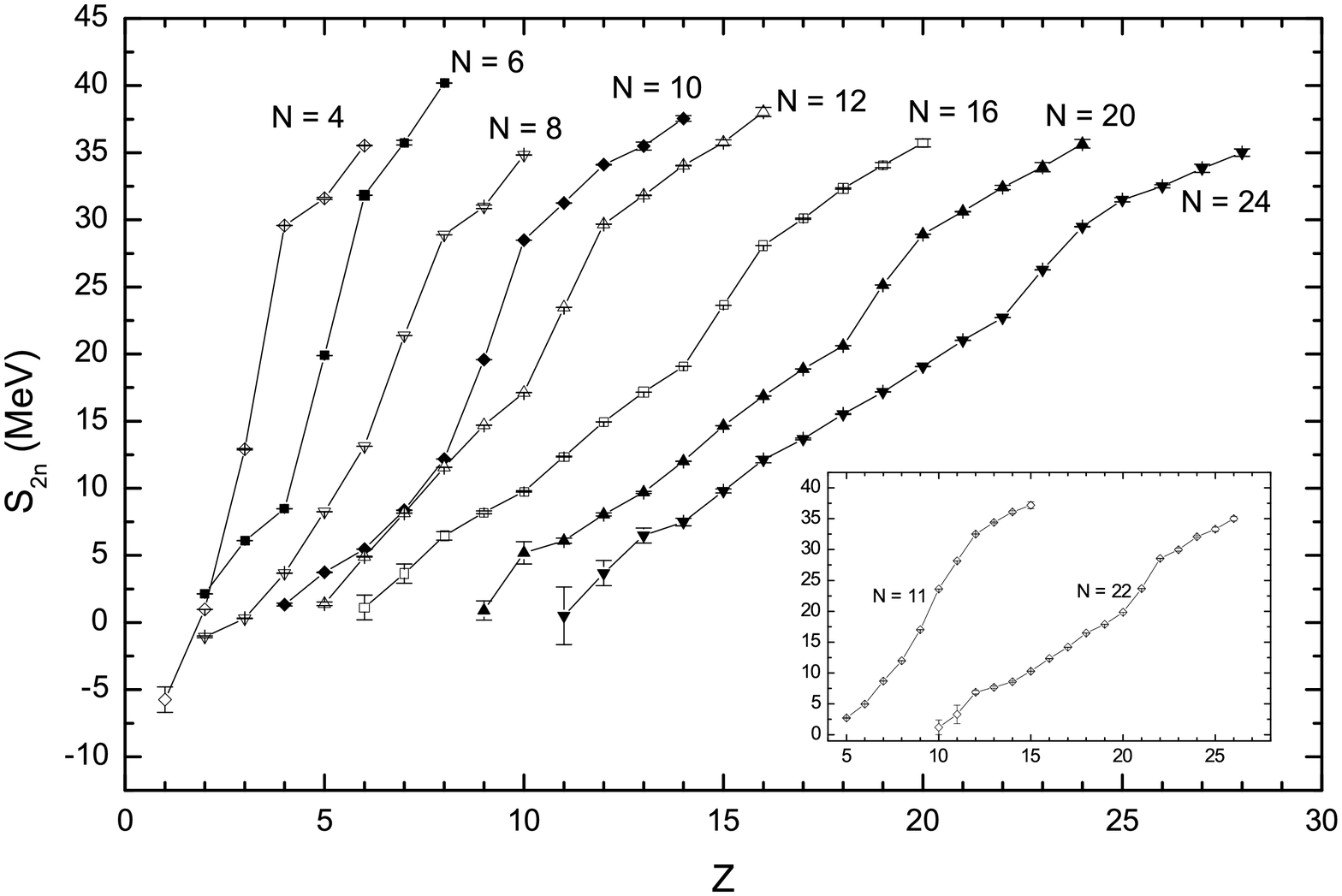}
\end{figure}

\newpage

\begin{figure}
\caption{Single neutron separation energy as a function of N for fixed Z}
\epsfclipon
\epsfxsize=0.99\textwidth
\epsfbox{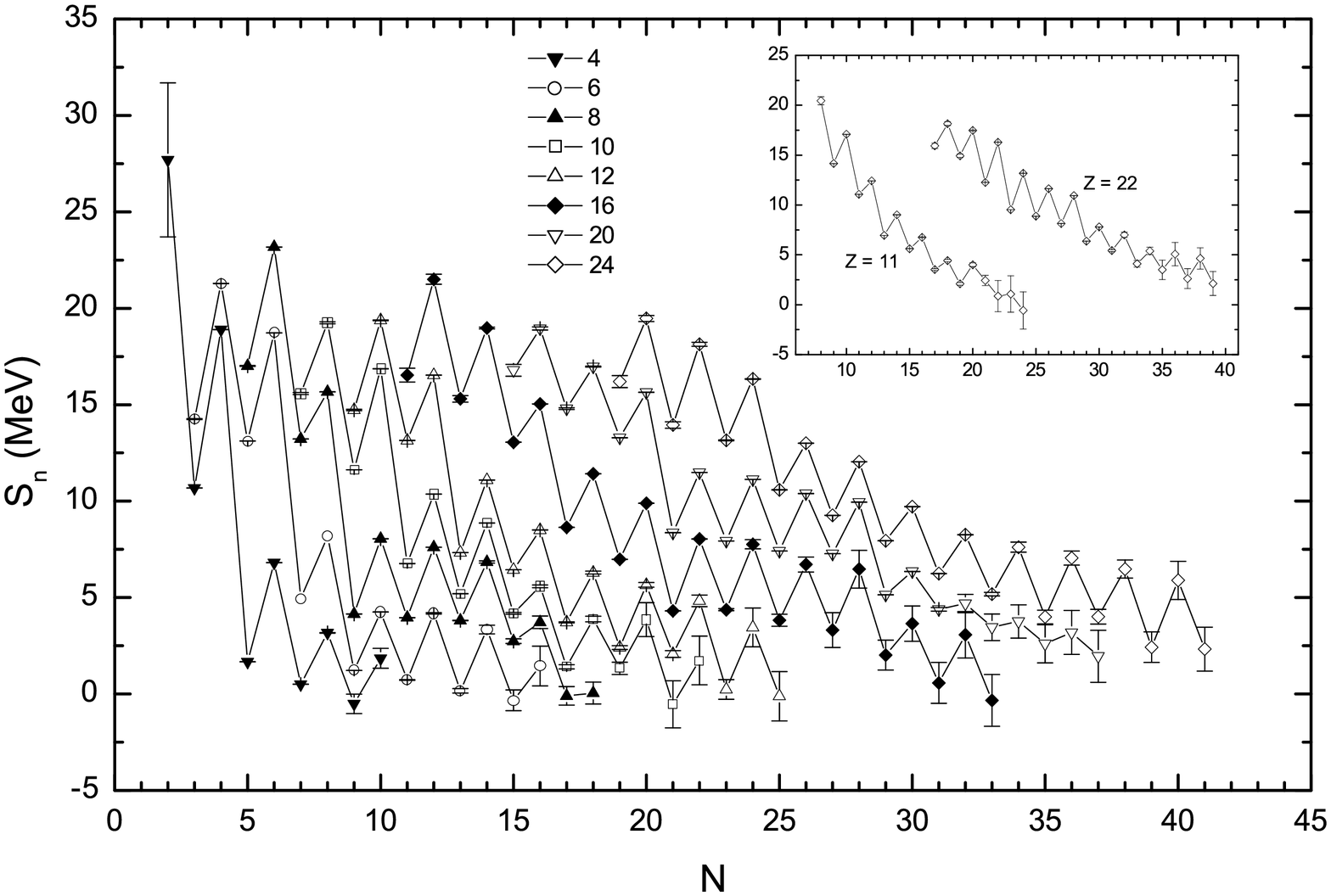}
\end{figure}

\newpage

\begin{figure}
\caption{Double neutron separation energy as a function of N for fixed Z}
\epsfclipon
\epsfxsize=0.99\textwidth
\epsfbox{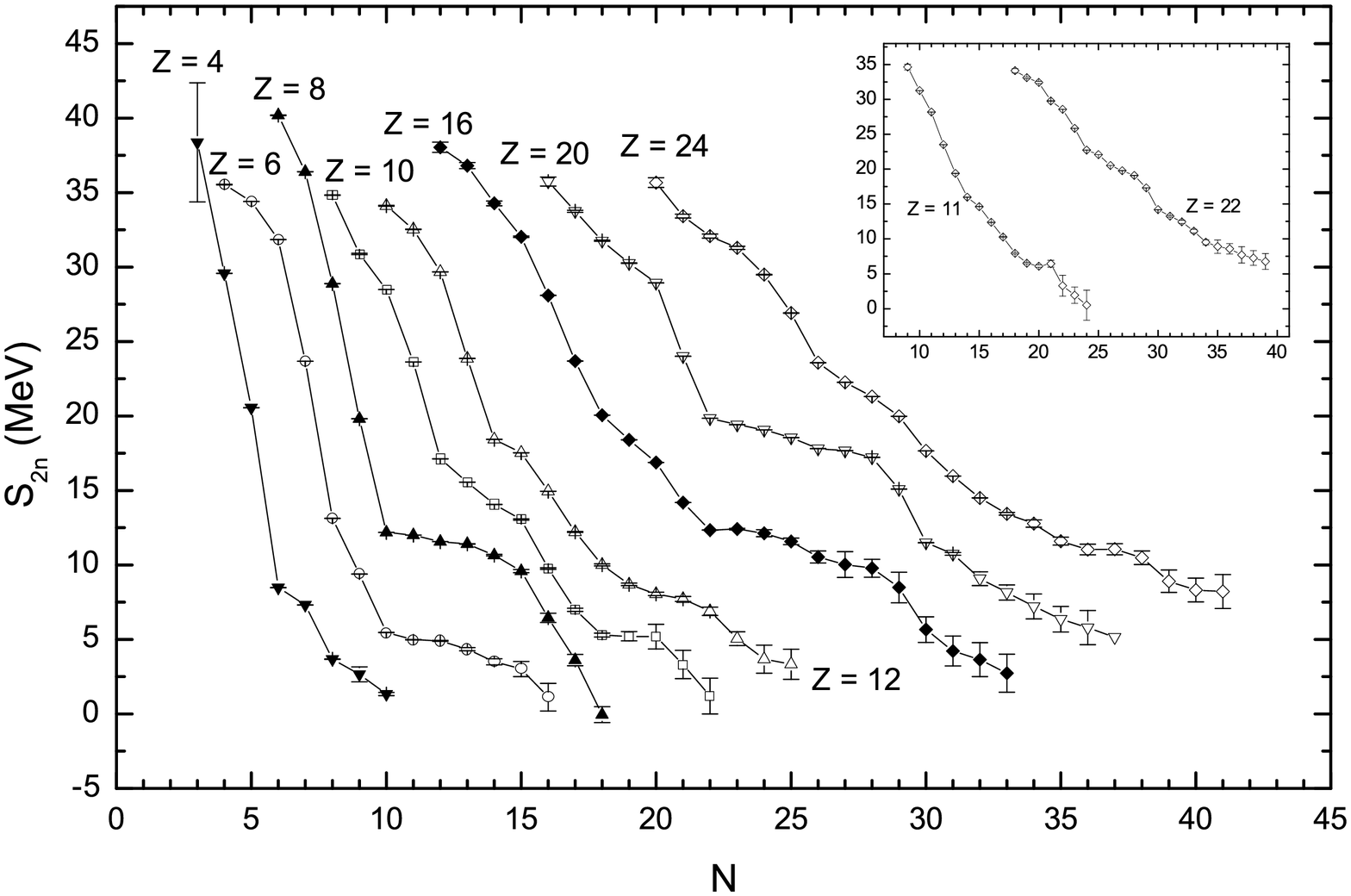}
\end{figure}

\newpage

\begin{figure}
\caption{Single proton separation energy as a function of Z for fixed N}
\epsfclipon
\epsfxsize=0.99\textwidth
\epsfbox{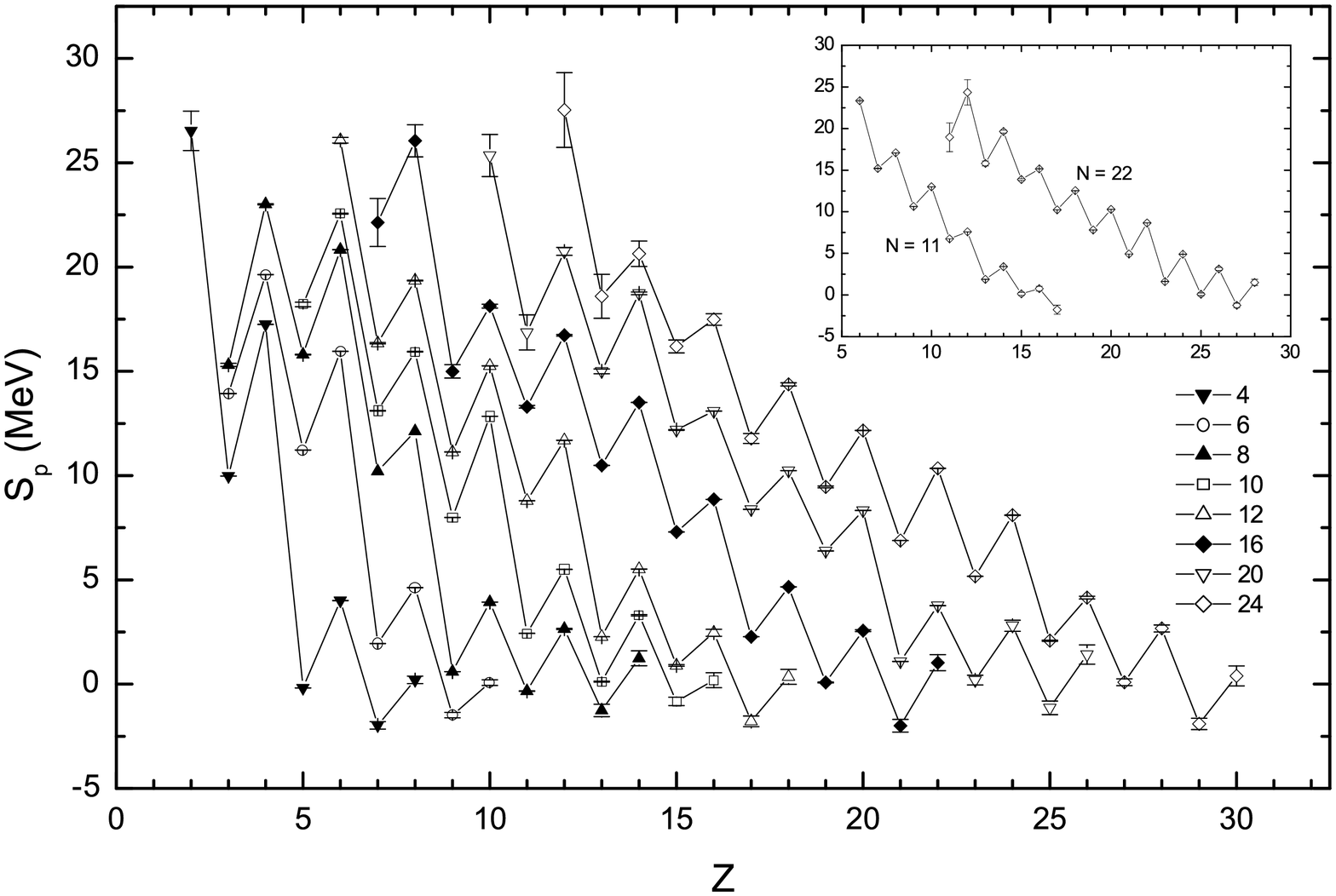}
\end{figure}

\newpage

\begin{figure}
\caption{Double proton separation energy as a function of Z for fixed N}
\epsfclipon
\epsfxsize=0.99\textwidth
\epsfbox{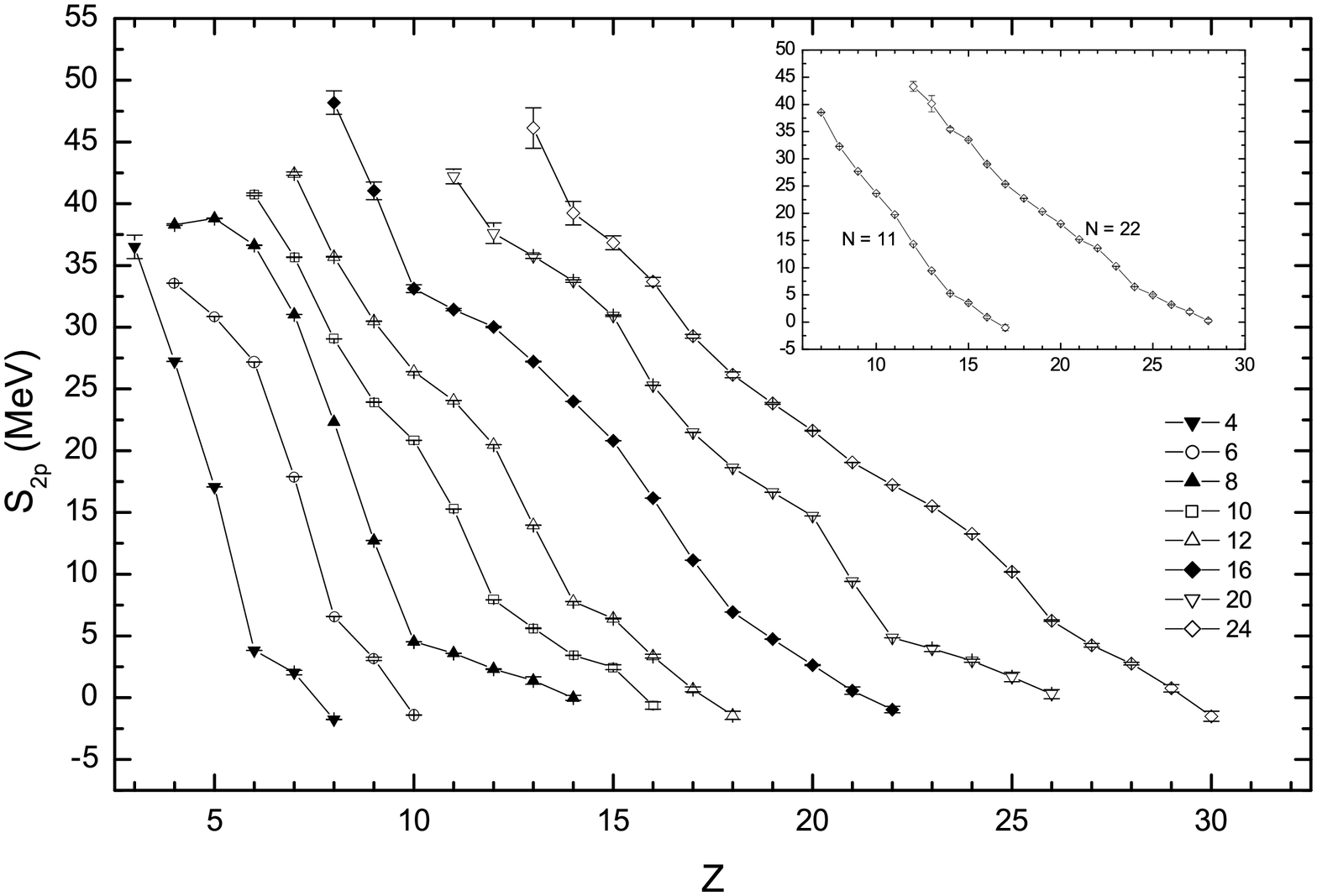}
\end{figure}

\newpage

\begin{figure}
\caption{Single proton separation energy as a function of N for fixed Z}
\epsfclipon
\epsfxsize=0.99\textwidth
\epsfbox{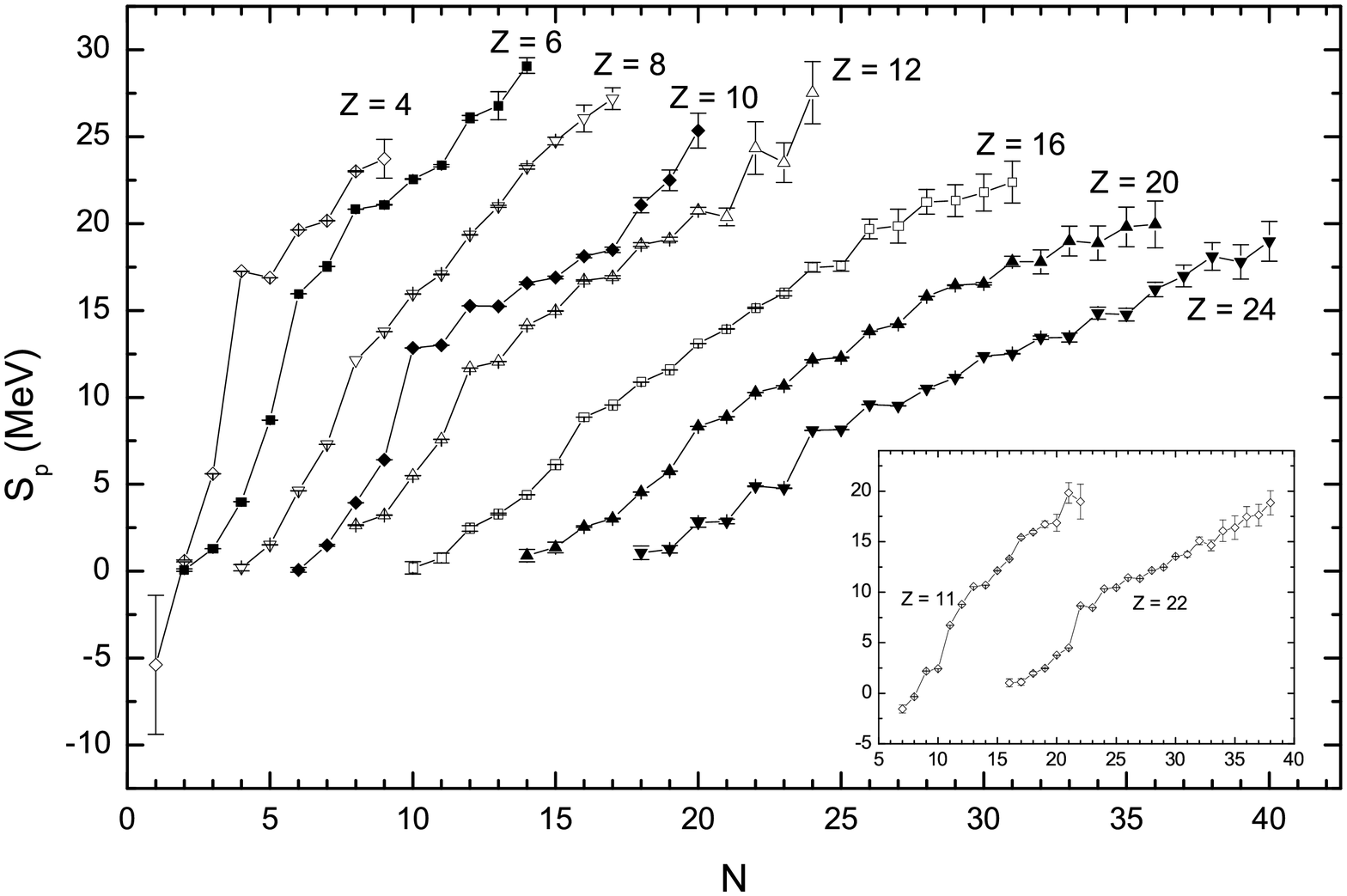}
\end{figure}

\newpage

\begin{figure}
\caption{Double proton separation energy as a function of N for fixed Z}
\epsfclipon
\epsfxsize=0.99\textwidth
\epsfbox{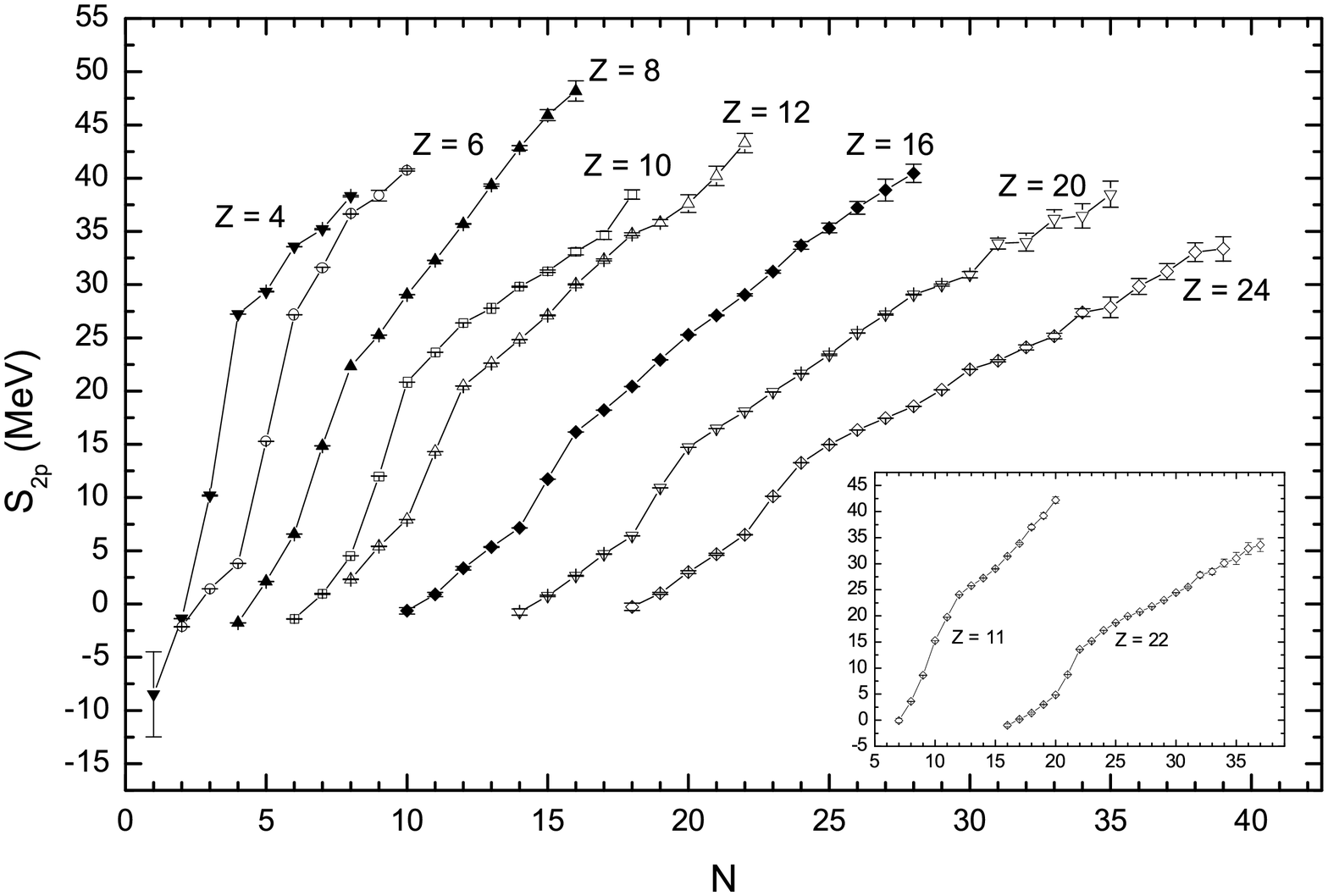}
\end{figure}

\end{document}